**Risk management for analytical methods: conciliating objectives of methods, validation phase and routine decision rules.**

M. Maumy (1), B. Boulanger (2), W. Dewe (2) A. Gilbert (3) & B. Govaerts(3)
(1) Université Louis Pasteur, Laboratoire de Statistique, Strasbourg
(2) Eli Lilly, European Early Phase Statistics, Belgique
(3) UCL, Inst. de Statistique, Belgique

**Introduction**

In the industries that involved either chemistry or biology, such as pharmaceutical industries, chemical industries or food industry, the analytical methods are the necessary eyes and hear of all the material produced or used. If the quality of an analytical method is doubtful, then the whole set of decision that will be based on those measures is questionable. For those reasons, being able to assess the quality of an analytical method is far more than a statistical challenge; it's a matter of ethic and good business practices. Many regulatory documents have been releases, primarily ICH and FDA documents in the pharmaceutical industry (FDA, 1995, 1997, 2001) to address that issue.

**Objective of an analytical method**

The objective of a good analytical method is to be able to quantify accurately each of the unknown quantities that the laboratory will have to determine. In other terms, what is expected from an analytical method is that the difference between a measurement (x) and the unknown "true value" ($\mu_T$) of the sample be small or inferior to an acceptance limit, i.e.:

$$-\lambda < x - \mu_T < \lambda \quad \Leftrightarrow \quad |x - \mu_T| < \lambda \qquad \text{Eq. 1}$$

With λ, the acceptance limit, which can be different depending on the requirements of the analyst or the objective of the analytical procedure. The objective is linked to the requirements usually admitted by the practice (e.g. 1% or 2% on bulk, 5% on pharmaceutical specialties, 15% for biological samples, 30% for ligand binding assays suc as RIA or ELISA, etc.).

A procedure is acceptable if it is very likely that the difference between each measurement (x) of a sample and its "true value" ($\mu_T$) is inside the acceptance limits [-λ,+λ] predefined by the analyst. The notion of "good analytical procedure" with a known risk can translate itself by the following equation (Boulanger et al. 2000a; 2000b):

$$P(|x - \mu_T| < \lambda) \geq \beta \qquad \text{Eq. 2}$$

With $\beta$ the probability a measure will fall inside the acceptance limits.

All analytical methods can be characterized by a "true bias" $\mu_M$ (systematic error), and a "true precision" $\sigma_M$ (random error). If the "true bias" $\mu_M$ and a "true precision" $\sigma_M$ are known, then assuming the measurements follow a normal distribution, the probability in Eq. 2 is easy to obtain using the normal distribution classically. This lead to define the Acceptance Region, i.e. the set of ($\mu_M, \sigma_M$) such that the probability of Eq. 2 is greater than $\beta$. Figure 1 shows, inside the curves, the Acceptance Region for various values of $\beta$ (99%, 95%, 90%, 80% and 66.7%) when acceptance limits are fixed to [-15%,+15%] as recommended by ICH (1995,1997) and FDA (2001) for bioanalytical methods. Logically, as it can be seen on Figure 1, the greater the variance of measure or greater the bias, the less likely a measure will fall within the acceptance limits.

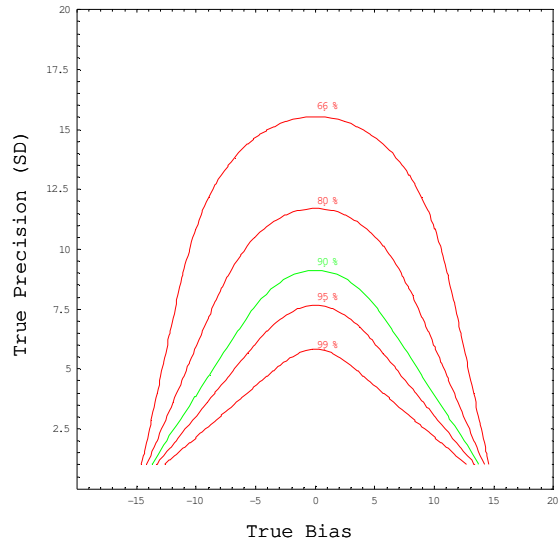

**Figure 1 :** Acceptance regions of analytical methods depending on the true bias and true precision for various probability to provide measures within the acceptance limits fixed in this example to [15%,+15%].

## Objective of pre-study validation experiments.

Before an analytical method can be used in routine for qualifying unknown samples, it's the practice to perform a more or less extensive set of experiments to evaluate if the analytical method will be able to achieve it's objective as stated above. Those experiments are usually called "pre-study validation" as opposed to the "in-study validation" experiments that required the use of QC samples used in routine and inserted in the unknown samples.

The "true bias" $\mu_M$ (systematic error), and a "true precision " $\sigma_M$ (random error) are intrinsic properties of all analytical procedures and are unknown. For that reason experiments are required before using the method in routine (pre-study validation) or QC are inserted among the unknown samples (in-study validation) to allow the user to obtain estimates of bias and of precision of the analytical method. These estimates of bias ($\hat{\mu}_M$) and variance ($\hat{\sigma}_M$) are intermediary but obligatory steps to evaluate if the analytical procedure is likely to provide accurate measure on each of the unknown quantities, i.e. to fulfill it's objective.

Consequently, the objective of the validation phase is to evaluate whether, given the estimates of bias $\hat{\mu}_M$ and variance $\hat{\sigma}_M$, the expected proportion of measures that will fall within the acceptance limits is greater than a predefined level of proportion, say β, as show in Eq. 3:

$$E_{\hat{\mu},\hat{\sigma}}\{P[|x_i - \mu_T| < \lambda]/\hat{\mu}_M, \hat{\sigma}_M\} \geq \beta \qquad \text{Eq. 3}$$

## Pre-study Decision rule

However there exist no exact solution for Eq. 3. An easy alternative to make a decision as already proposed by other authors (Boulanger et al. 2000; Hubert et al.

2004; Hoffman & Kringle, 2005) is to compute the β-expectation tolerance interval (Mee, 1984):

$$E_{\hat{\mu}_M,\hat{\sigma}_M}\{P_X[\hat{\mu}_M - k\hat{\sigma}_M < X < \hat{\mu}_M + k\hat{\sigma}_M / \hat{\mu}_M,\hat{\sigma}_M]\} = \beta \qquad \text{Eq. 4}$$

where the factor k is determined so that the expected proportion of the population falling within the interval is equal to β. If the β-expectation tolerance interval obtained Eq. 4 is totally included within the acceptance limits [-λ,+λ], i.e. if ($\hat{\mu}_M - k\hat{\sigma}_M > -\lambda$ and $\hat{\mu}_M + k\hat{\sigma}_M < +\lambda$) then the expected proportion of measurements within the same acceptance limits is greater or equal to β, i.e. Eq. 3 is also verified in that case. Note that the opposite statement is not true, i.e. if either $\hat{\mu}_M - k\hat{\sigma}_M < -\lambda$ or $\hat{\mu}_M + k\hat{\sigma}_M > +\lambda$ doesn't imply that the expected proportion is smaller than β.

**Conciliating pre-study and in-study decision rules**

The central question what value for β to use in Eq. 1, Eq. 2 and Eq. 4 to guarantee compliance with regulatory documents and ensure that the pre-study rule (Eq. 4) will be in accordance with the in-study acceptance criteria. An in-study rule that is however largely admitted in the community, also called "4-6-15" rule and defined in the FDA guidance [FDA, 2001]is : "*At least four of every six QC samples should be within "15% of their respective nominal value. Two of the six QC samples may be outside the "15% of their respective nominal value, but not both at the same concentration.*"

This rule contains two important information: the first –explicit- that the acceptance limits for bioanalytical methods are [-15%,+15%] and the second –implicit- the probability that a measure should fall within those acceptance limits. All will agree that the "4-6-15" rule should be accepted often, for example in 90% of cases, otherwise it become economically counterproductive to maintain an analytical method that frequently lead to reject runs. Then if one agree that a good analytical method is the one that make the "4-6-15" rule being accepted at least in 90% of the cases, then β is the probability of success that lead to obtain at least 4 successes out of 6 trials. Inverting the Binomial distribution is the solution of this problem. It can be show, but would be too long to demonstrate here, that to have the "4-6-15" rule being accepted for 90% of the run, then the probability success should be equal to 80%. This can be seen on Figure 2 that shows the probability of accepting the 4-6-15 rule as a function of the probability that a measure will fall within the acceptance limits [-15%,15%]. This contrasts with the proposal frequently encountered [DeSilva, 2003; Hoffman & Kringle 2005] that 4/6 or 66.7% of the measures must lie within the acceptance limits. Proposing 66.7% as value for β as suggested by those authors can lead to reject up to 32% of the runs (!) as can be seen on Figure 2. This is certainly not the intend of the authors and rather results from an erroneous interpretation of the Binomial distribution.

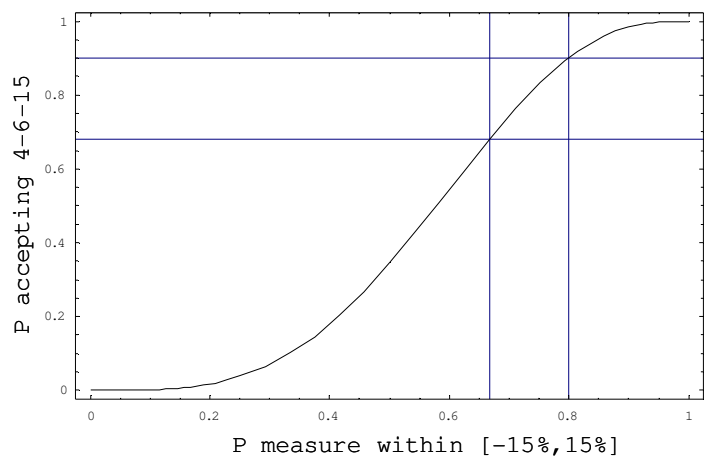

**Figure 2 :** Probability of accepting the 4-6-15 rule [P(Y≥4)] as a function of the probability that a measure will fall within the acceptance limits [-15%,15%] i.e. assuming Y ~ Bi(6,p).

## Conclusion:

The expected proportion β of measure that should fall within the acceptance limits [-λ,λ] must be at least equal to 80% to guarantee that at least 90% of the runs will be accepted with the 4-6-λ rule. Taking 0.8 as value for Eq. 2, 3 and 4 allows making pre-study and in-study decision rules consistent. But as seen on Figure 2, the 4-6-15 rule lacks of power. For example if there is only 50% chance that a measure lies within the acceptance limits, the runs will be accepted in about 35% of the cases. 35% too high from a consumer perspective. The only solution is to improve the in-study rule, using for example a 5-6-15 or even a 10-12-15 rule and to adapt the β value accordingly for the pre-study validation. This will be part of future works.